\documentclass[12pt]{article}

\def\ba{\begin{eqnarray}}
\def\ea{\end{eqnarray}}
\def\be{\begin{equation}}
\def\ee{\end{equation}}

\def\NSNS
\def\RR{R$\otimes$R }
\def\NSNS{NS$\otimes$NS }
\def\npb#1(#2)#3     {Nucl. Phys. {\bf B#1} (#2) #3 }
\def\xxx#1           {hep-th/#1 }
\def\det{\hbox{\rm det}}

\def\Zop{\bbbz}

\def\Nop{\bbbn}
\def\bbbz {{\sf Z\!\!Z}}

\def\bbbn {{\rm I\!N}}
\def\pmb#1{\setbox0=\hbox{#1}
 \kern-.025em\copy0\kern-\wd0
 \kern.05em\copy0\kern-\wd0
 \kern-.025em\raise.0433em\box0 }

\def\I{{\cal I}}

\def\3{\ss}
\def\sq{\hbox{\rlap{$\sqcap$}$\sqcup$}}
\def\qed{\ifmmode\sq\else{\unskip\nobreak\hfil
\penalty50\hskip1em\null\nobreak\hfil\sq
\parfillskip=0pt\finalhyphendemerits=0\endgraf}\fi}
\def\half {\frac{1}{2}}

\newcommand{\ket}[1]{|#1\rangle}
\newcommand{\bra}[1]{\langle#1|}

\begin{document}

\thispagestyle{empty}
\def\thefootnote{\fnsymbol{footnote}}
\begin{flushright}
  hep-th/9910217\\
  DAMTP-1999-154 
 \end{flushright}  
\vskip 0.5cm

\begin{center}\LARGE
{\bf Non-BPS States and Heterotic -- Type I$^\prime$ Duality}
\end{center}
\vskip 1.0cm
\begin{center}
{\centerline{  Tathagata Dasgupta, Bogdan Stefa\'nski, jr.\footnote{T.Dasgupta,
 B.Stefanski@damtp.cam.ac.uk}}}

\vskip 0.5 cm
{\it Department of Applied Mathematics and Theoretical Physics\\
University of Cambridge, Silver Street, \\
Cambridge, CB3 9EW, U.K.}
\end{center} 

\vskip 1.0cm

\begin{center}
October 1999
\end{center}

\vskip 1.0cm

\begin{abstract}
There are two families of non-BPS bi-spinors in the
perturbative spectrum of the nine dimensional heterotic string charged under
the gauge group $SO(16)\times SO(16)$. The relation between these
perturbative non-BPS states and certain non-perturbative non-BPS D-brane
states of the dual type I$^\prime$ theory is exhibited. The relevant 
branes include a $\Zop_2$ charged non-BPS D-string, and a bound state of such a
D-string with a fundamental string. The domains of stability of these
states as well as their decay products in
both theories are determined and shown to agree with the
duality map.   

\end{abstract}

\vfill
\setcounter{footnote}{0}
\def\thefootnote{\arabic{footnote}}
\newpage

\section{Introduction}
\setcounter{equation}{0}

Over the past couple of years stable non-BPS states and D-branes have
opened a new direction to our understanding of string theory. Reviews
of these developments can be found in~\cite{sen,lr,sch}. Two approaches
have been used to construct and analyse non-BPS D-branes. In the
first approach~\cite{sen1,sen2,sen3,sen4,sen5} 
non-BPS D-branes are constructed by
tachyon condensation as bound states of brane-anti-brane
pairs. This construction permits for a classification of D-brane
charges in terms of K-theory~\cite{wittK}. The second
approach uses the boundary state formalism~\cite{polcai,clny1,clny,li,gg}, 
to describe D-branes as coherent states in the closed
string theory satisfying a number of consistency 
conditions~\cite{bg1,bg2,gsen,gs}. 
Since this latter approach provides an explicit
boundary conformal field theory description of non-BPS D-branes, we use 
this second approach. 

$SO(32)$ heterotic string theory is conjectured to be
non-perturbatively dual to type I string theory in ten dimensions
\cite{wittD}.  
It should therefore be possible to identify suitable perturbative
non-BPS states of the heterotic string with non-BPS D-brane states in the
type I theory. The most familiar example of stable
non-BPS states are the states transforming in the spinor
representation of the gauge group of the $SO(32)$ heterotic
string theory. They arise in the first excited level and are
absolutely stable due to charge conservation but are not BPS as $N=1$
supersymmetry algebra has no central charges. 
The dual state in the type I theory is a
stable $\Zop_2$-valued non-BPS D-particle~\cite{sen4,wittK}, which
can be described as a tachyonic kink solution on the
D1-$\overline{\mbox{D}1}$ pair \cite{sen2}. 

In this paper we test the S-duality between the heterotic string
theory on $S^1$ with gauge group $SO(16) \times SO(16)$ and the type
I$^\prime$ theory on $S^1$ with the same gauge group. In such a
configuration the two heterotic theories are T-dual to each
other~\cite{nsw,gins}. The type I$^\prime$ theory is
an orientifold of type IIA by $\I_9\Omega$, where $\I_9$
reverses the sign of $x^9$ and $\Omega$ is the
world-sheet parity operator. This orientifold can be thought of as
two O8-planes~\cite{t1orb} 
at $x^9=0$ and $x^9=\pi R_9$, which in the $SO(16)\times
SO(16)$ point in moduli space has eight D8-branes, and their images placed on
each of the O8-planes to cancel the tadpole locally.\footnote{For a
non-technical review of $\Zop_2$ orientifolds of type II theories see
for example~\cite{mukhi}.} The positions 
of the D8-branes on the interval correspond in the T-dual
type I theory to a Wilson line. 

The conserved charges of the type I string theory are
Kaluza-Klein (KK) momentum and D-string winding number, and those of
type I$^\prime$ are winding and D-particle numbers. The duality map
relating various parameters
between type I$^\prime$ and heterotic theory is given by~\cite{pol-witt} 
\be
R_h = {1 \over \sqrt{R_{I'}\lambda_{I'}}}, \,\,\,\,\, \lambda_h =
{R_{I'} \over \lambda_{I'}}, \,\,\,\,\, G^h_{MN} = G^{I'}_{MN} {R_{I'}
\over \lambda_{I'}},   \label{dmap}
\ee
where $G_{MN}$ is the nine-dimensional metric ($M,N = 0, \ldots ,8$),
$\lambda_h$ and $\lambda_{I'}$ are the heterotic and type I$^\prime$ coupling
constants, and $R_h$ and $R_{I'}$ are the radii of the circle when
measured in the heterotic and type I$^\prime$ metric, respectively. From 
this relations one can see that the heterotic KK momentum is mapped to
type I$^\prime$ winding, and heterotic winding is mapped to the type I$^\prime$
D-particle number.       
  
Under the unbroken $SO(16)\times SO(16)$ gauge group 
the states in the spinor conjugacy class of $SO(32)$ representations
decompose into two sets of bi-spinors, denoted ${\cal A}$ and ${\cal
B}$. In the type I$^\prime$ theory, ${\cal A}$ are shown to correspond
to a $\Zop_2$-valued non-BPS D-string stretching along the 
interval which is T-dual to the $\Zop_2$-valued
D-particle on $S^1$ in type I theory. We show that these
bi-spinors are unstable against a decay into two single spinor states
for radii less than a critical radius $R_h$. In the type I$^\prime$ theory, 
this corresponds to the appearance of a tachyon in the
open string spectrum with endpoints on the $\Zop_2$-valued
D-string. For radii greater than a certain critical
radius $R_{I'}$, the D-string decays into
a D0-brane at one O8-plane and an anti-D0-brane on the other
O8-plane. The  
$\Zop_2$ charge of the decaying D-string is encoded in the $\Zop_2$
choice of locations for the decay products. We determine the masses of
the various states and show that at the critical radius they are equal
to one another, indicating that the deformation is marginal. 
Further we show that the critical radii $R_h$
and $R_{I'}$ agree with (1.1) qualitatively. This provides a test of
the type I$^\prime$-heterotic duality beyond the constraints of BPS
states. 

The second class of bi-spinors ${\cal B}$
can be thought of as bound states of the ${\cal A}$ bi-spinors 
and certain bi-vectors. In
type I$^\prime$ the ${\cal B}$ bi-spinors correspond to a bound
state of a $\Zop_2$-valued non-BPS D-string and F-string both
stretching along the interval. We 
show that such a bound state does indeed exist in type I$^\prime$.
Although
the ${\cal B}$ bi-spinor (the (F,D) bound state) is stable against
decay into an ${\cal A}$ bi-spinor (the non-BPS D-string) and the
bi-vector (the fundamental string) as we will see, it is
not always stable against decay into the two different single spinor
states (a D0-$\overline{\mbox{D}0}$ pair) and a bi-vector (a fundamental
string stretching along $x^9$). The mass of the bound state and
domains of stability of the heterotic and type I$^\prime$ states are
computed. The regimes of stability in the two theories are shown to be 
qualitatively the same. In the
T-dual picture, the (F,D) bound state corresponds 
to a $\Zop_2$-valued D-particle with a constant velocity along
$S^1$ as the effect of adding a fundamental string
to the D-string of type I$^\prime$ is to give the type I D-particle a
KK momentum. We show that, unlike in the previous case,
presently the non-BPS mass of the (F,D) bound state does not
match with its decay product at the critical radius. This indicates
that the transition from the non-BPS bound state to the D-particle
pair and a F-string is not a marginal deformation. 

The paper is organised as follows. In section 2 we briefly review the heterotic
string on $S^1$ and point out that there are two kind of non-BPS
bi-spinors arising from the unbroken gauge group $SO(16) \times
SO(16)$. The duality map
(1.1) is tested by comparing the masses of certain BPS states, namely
bulk D-particle and fractional D-particle and anti-D-particle,
in section 3. In section 4 the non-BPS bi-spinor states are
analysed. In section 5 we give the type I$^\prime$ analysis following
boundary state approach. We conclude and raise some open problems in
section 6.
\section{Review of Heterotic String on $S^1$}\label{sec2}
\setcounter{equation}{0}
The left- and right-moving momenta of an $SO(32)$ heterotic string 
compactified on $S^1$ are given by \cite{gins}\footnote{Throughout the
paper we follow the convention $\alpha'_h = 2,\,\alpha'_{I'} = 1$.} 
\begin{eqnarray}
\left( p_L \,|\, p_R \right) & = & \left( V + Aw_h \; , \; {p_h \over
R_h} + {w_h R_h \over 
2}\,\,\, | \,\,\, {p_h \over R_h} - {w_h R_h \over 2}\right)\,,
\end{eqnarray}
where $V$ is an element of the internal $\Gamma^{16}$ lattice,
$R_h$ and $p_h$ are the compactification radius and physical momentum,
respectively, while $w_h \in {\sf Z\!\!Z}$ is the winding number
respectively. In terms of the background gauge field, $A$ (or Wilson
line) $p_h$ is given by 
\be
p_h = n_h - V \cdot A - w_{h}A^2/2\,,
\ee 
where $n_h \in \Zop$ denotes the Kaluza-Klein (KK)
momentum. Physical states satisfy the level matching condition
\be
p_L^2 + 2 \left( N_L - 1 \right) =   
p_R^2 + 2 \left( N_R - c_R \right)\,, \label{lmatch}
\ee
where $N_L$ and $N_R$ are left- and right-moving excitation numbers,
and $c_R = 0$ and $1/2$ for the right-moving fermions in the periodic
(R) and anti-periodic (NS) sectors, respectively. For BPS states all
the right moving oscillators should be in the ground state: 
$N_R = c_R$~\cite{dabh-harv}. The heterotic mass formula is given by 
\be
m_h^2  =  {1 \over 2} p_L^2 + \left( N_L - 1 \right) + 
{1 \over 2} p_R^2 + \left( N_R - c_R \right)\,,
\ee
which using the level-matching condition~(\ref{lmatch}) becomes
\be
m_h^2  =  p_R^2 + 2 \left( N_R - c_R \right).
\ee
The states with $N_L = 1$ and $V^2 = 0$ are KK excitations of either
the gravity multiplet or one of the vector multiplets associated with the
Cartan subalgebra of the gauge group. But there are additional
massless states having $N_L = 0$ and $p_L^2 = 2$. In the zero
winding sector ($w_h = 0$) we have $V^2 = 2$ and $p_h = 0$, hence
states (V) are roots of $SO(32)$. For a non-trivial Wilson line $A$, these
states correspond to the roots of the unbroken subgroup of
$SO(32)$. Without a
Wilson line, the heterotic string has a $R \rightarrow 1/R$ symmetry
giving rise to an enhanced $SU(2)$ gauge symmetry at the self-dual
point. More generally, for a 
nontrivial Wilson line, $A$, the winding sectors give additional
massless states at some critical radius. In this paper we consider only 
the Wilson line, $A = (0^8 , (1/2)^8)$, which corresponds to an 
$SO(16) \times SO(16)$ gauge group and critical radius zero. 
In type I$^\prime$ theory this is equivalent to placing eight 
D8-branes with mirrors at each O8-plane giving rise to a constant
dilaton and metric background in the bulk~\cite{pol-witt,bgl}.

The $2^{n-1}$-dimensional spinor representation is the smallest 
representation of the ten dimensional spinor conjugacy class of
$SO(32)$. In nine dimensions, under the Wilson line $A=(0^8;({1 \over
2})^8)$ with unbroken gauge group $SO(16)
\times SO(16)$, this decomposes as 
\be
2^{15}  \longrightarrow  ( 2^{7} , 2^{7} ) \oplus ((2^{7})',(2^{7})')\,,
\ee
giving two kinds of bi-spinors with 
\ba
({\cal A}):\,\,\,\,\,\,\,\left( \underbrace{ \pm {1 \over
2} , \ldots ,  
\pm {1 \over 2}}_{\mbox{even no. of `+'}} ; \underbrace{ \pm 
{1 \over 2} , 
\ldots , \pm {1 \over 2}}_{\mbox{even no. of `+'}} \right) & \in & 
( 2^7 , 2^7 ),  \nonumber \\ 
\left({\cal B}\right):\,\,\,\,\,\,\,\left( \underbrace{ \pm {1 \over
2} , \ldots ,  
\pm {1 \over 2}}_{\mbox{odd no. of `+'}} ; \underbrace{ \pm
{1 \over 2} , 
\ldots , \pm {1 \over 2}}_{\mbox{odd no. of `+'}} \right) & \in &
((2^{7})' ,(2^{7})')\,.  
\ea
In later sections we will show that in type I$^\prime$ theory, the first
set of bi-spinors 
corresponds to a ${\sf Z\!\!Z}_2$-charged non-BPS D-string stretching
along the interval with end-points  at the two O8-planes. The second set
corresponds to a {\it bound state} of the
${\sf Z\!\!Z}_2$-charged 
non-BPS D-string with an F-string, both stretching along the interval.
\section{Single Spinor States}\label{sec3}
\setcounter{equation}{0}
In this section we describe heterotic single spinor states that are charged
under one of the $SO(16)$'s. In type I$^\prime$, they turn out to be
fractional D-particles (or D0$_f$) stuck on an
O8-plane. Their masses and bulk RR charges are half of those of the
bulk D-particles. 

A bulk D-particle is dual to a heterotic string with non-trivial
winding, $w_h = 2$ and the physical momentum, $p_h =
0$~. As $V+Aw_h= 0$ with $V = \left( 0^8 ; (-1)^8 \right)$,
these states are not charged under either of the $SO(16)$'s. The 
level-matching condition implies that $N_R = c_R$ and $N_L = 1$. Hence
these are BPS states and are KK excitations of either the gravity
multiplet or the vector multiplets
in the Cartan subalgebra. The heterotic mass formula gives
\be
m_h \left( D0_{\mbox{\scriptsize bulk}} \right) = R_h\,.
\ee
Using~(\ref{dmap}) the corresponding type I$^\prime$ bulk
D-particle mass turns out to be
\be 
m_{I'} \left(\mbox{D}0_{\mbox{\scriptsize bulk}} \right) 
= \lambda_h^{1/2} m_h \left(\mbox{D}0_{\mbox{\scriptsize bulk}}
\right) = {1 \over \lambda_{I'}}\,.
\ee
Consider next the single spinor states 
\be
V_1  =  \left( \underbrace{ \left(\pm {1 \over 2} \right)^8
}_{\mbox{even no. of `+'}} ; 0^8 \right) ,\;\;\;
V_2  =  \left( 0^8 ; \underbrace{ \left(\pm {1 \over 2} \right)^8
}_{\mbox{even no. of `+'}} \right)\,. \label{v1v2}
\ee 
where $V_1$ is of the form $V + Aw_h$, with $V = (( \pm {1\over 2})^8 ; 
( - {1 \over 2})^8) \in \Gamma^{16}$, with an even
number of `+' signs, $w_h=1$ and vanishing physical momentum: 
$p_h = 0$ (for KK
momentum, $n_h = -1$). The type I$^\prime$ D-particle number, 
is given by \cite{bgl} $n_{I'} = w_h/2 = 1/2$. The level-matching 
condition gives $N_L = 0$ 
and $N_R = c_R$, thus $V_1$ is a BPS state whose mass is 
\be
m_h \left(V_1 \right) = {R_h \over 2}\,.
\ee
Using the duality relations~(\ref{dmap}), the type I$^\prime$ mass is
\be
m_{I'} \left(V_1 \right) = {1 \over 2 \lambda_{I'}}\,, \label{mdf1} 
\ee
which is half of the D-particle's mass. This 
shows that $V_1$ corresponds to a fractional D-particle
stuck on the $x^9=0$ O8-plane. 

Similarly $V_2$ is of the form $V + Aw_h$ where $V = \left( 0^8 ;
1^{2n} , 0^{8-2n} \right) \in \Gamma^{16}$ with $w_h = -1$, and
$n=0,\dots,4$.  
The level-matching condition implies that the state is BPS ($N_R = 
c_R$), with vanishing physical momentum, $p_h = 0$ (i.e. KK momentum,
$n_h = n-1$). The mass formula gives 
\be
m_h \left( V_2 \right) = {R_h \over 2}\,,
\ee   
which in type I$^\prime$ units is
\be
m_{I'} \left(V_2\right) = {1 \over 2 \lambda_{I'}}\,.
\label{mdf2} 
\ee   
This shows that $V_2$ corresponds to a fractional anti-D-particle
stuck at the $x^9=\pi R_9$ O8-plane.
\section{Bi-spinor States}
\setcounter{equation}{0}
In this section we discuss in some detail the bi-spinors ${\cal A}$
and ${\cal B}$. In particular we demonstrate that they are non-BPS
states, which are stable in certain regions of the moduli space.
\subsection{Bi-spinors in $\cal A$}
The lightest states in $\cal A$ come from the zero winding
sector. Since $A \cdot V \in {\sf Z\!\!Z}$ we choose $p_h = 0$. 
For $N_L = 0$ the level matching condition implies that the states are
non-BPS: $N_R = 
1 + c_R$. The mass of the states with trivial winding,
vanishing momentum, $N_L = 0$ and $N_R = 1 + c_R$ is
\be
m_h \left( V \in \left( {\bf 128 , 128} \right) \right) = \sqrt 2 \,. 
\ee
Note that there could be BPS states satisfying $N_R = c_R$ with $w_h =
\pm 1, \, p_h = 0$. But for such winding numbers the modified
lattice vector $\left( V + Aw_h \right)$, does not belong to $\Gamma^{16}$.

The bi-spinors ${\cal A}$ are charged under both the $SO(16)$'s. In
fact these non-BPS states have the same charges as $V_1$ and $V_2$,
the single spinor states discussed in section~\ref{sec3}. This
suggests that a decay process into the BPS single spinor states is
possible. Since the mass of the single
spinor states is radius dependent the decay too will depend on the
radius. As the mass of the single
spinor states is $R_h/2$, the bi-spinor states ${\cal A}$
decay into the single spinor states $V_1$ and
$V_2$ for
\be
R_h < \sqrt 2\,. \label{Rchet1}
\ee
In the following section we construct 
a D-brane state which corresponds to ${\cal A}$. This
 turns out to be a $\Zop_2$ D-string stretching 
along the interval. This D-string is the T-dual of the $\Zop_2$ 
D-particle of type I. 
\subsection{Bi-spinors in $\cal B$}
The states, $V' \in \left( {\bf 128' , 128'} \right)$ (i.e. in sector
$\cal B$) have an odd number of $+{1 \over
2}$'s in both the first and second eight entries. Hence they can be
expressed as sum of a state, $V \in \left( {\bf 128 , 128} \right)$
(i.e. in sector $\cal A$) and a bi-vector state, $V_{bv} \in \left(
{\bf 16, 16} \right)$:     
\be
\left( \underbrace{ \pm {1 \over 2} , \ldots , 
\pm {1 \over 2}}_{\mbox{odd no. of `+'}} ; \underbrace{ \pm 
{1 \over 2} , 
\ldots , \pm {1 \over 2}}_{\mbox{odd no. of `+'}} \right)
\,\,=\,\, 
\left( \underbrace{ \pm {1 \over 2} , \ldots , 
\pm {1 \over 2}}_{\mbox{even no. of `+'}} ; \underbrace{ \pm 
{1 \over 2} , 
\ldots , \pm {1 \over 2}}_{\mbox{even no. of `+'}} \right)
\pm
\left( 1, 0^7 ; \pm 1, 0^7 \right)\,. \label{bsbv}
\ee
In type I$^\prime$ the state $\pm \left( 1, 0^7 ; \pm 1, 0^7
\right)$ corresponds to a fundamental string stretching along the interval.
The overall sign corresponds to the orientation of the string. 

The lightest states in $\cal B$ are those with $w_h = 0$. As $A \cdot
V \in {\sf Z\!\!Z} + 1/2$, these
have $p_h = \pm 1/2$. For $N_L = 0$ the level
matching condition implies that the states are
non-BPS: $N_R = 
1 + c_R$. Here $2n+1$ ($n = 0, \ldots ,3$) is the number of $+{1 \over
2}$'s in the last eight entries of $V \in \left( {\bf 128, 128}
\right)$. The mass formula 
for such states is
\be
m_h \left(V' \in \left( {\bf 128', 128'} \right) \right) = \left( {1 
\over 4R_h^2} + 2 \right)^{1/2} \,.  
\ee
On the other hand, the bi-vector states, $V_{bv}$ with $w_h = 0$,
$N_L = 0$ are BPS ($N_R = c_R$). The mass of these BPS states with
physical momentum, $p_h = \pm 1/2$ (as $A \cdot V = \pm 1/2$) is given
by 
\be
m_h \left( V_{bv} \in \left( {\bf 16, 16} \right)\right) = {1 \over
2R_h} \,. \label{mbv}
\ee
Since $m_h^2 \left( V \in \left( {\bf 128, 128} \right)\right) = 2$, we have
\be
m_h \left( V' \in \left( {\bf 128', 128'} \right)\right) < m_h \left(
V \in \left( {\bf 128, 128} \right)\right) + m_h \left( V_{bv} \in
\left( {\bf 16, 16} \right)\right)\,,  \label{hetin} 
\ee 
for a finite radius. Hence the states in
$\cal B$ are bound states of states in $\cal A$ and the bi-vector
states~(\ref{bsbv}). Although the bi-spinor, $\cal B$ is stable
against a decay into the bi-spinor $\cal A$ and
a bi-vector state, it is not always stable against decay into the 
single spinor states, $V_1$ and $V_2$, and a
bi-vector state. In particular this bi-vector is unstable for
\be
R_h < 1 \,. \label{Rchet2}
\ee
In the next section we show that in type I$^\prime$ theory the
bi-spinor ${\cal B}$ 
corresponds to a {\it bound
state} of a non-BPS $\Zop_2$ D-string with an F-string, both
stretching along the
interval. Equivalently, the states in $\cal B$ describe a non-BPS
D-string with a constant electric field on its world-sheet
\cite{witt1}. This ${\sf Z\!\!Z}_2$ D-string with constant
electric field becomes under T-duality, a ${\sf Z\!\!Z}_2$
D-particle with constant velocity along $S^1$~\cite{bachas,notesond}.
\section{Type I$^\prime$ analysis}\label{sec4}
\setcounter{equation}{0}
The states in $\cal B$ describe a non-BPS D-string
with a constant electric field on its world-sheet. Such a bound state
will exist as long as the constant electric field on
a $\Zop_2$ D-string is invariant under $\Omega \I_9$. 
If the gauge field is invariant under $\Omega \I_9$ the (F,D) bound state
can be stabilised from an unstable non-BPS D-string with constant
electric flux in type IIA by orientifolding. 
The NSNS $B$ field equation of motion has ${\cal F} = F + B$ 
as a source term~\cite{clny,witt1,li,schmid,gg} where 
$F = dA$ is the gauge field strength. Recall that in type I,
the gauge field vertex operator has tangential derivatives of the form
$dX^i/d\tau$ which are odd under $\Omega$, while the transverse
scalars have normal derivatives. As a result the gauge field is
projected out while  the transverse scalars survive.\footnote{In fact
a $\Zop_2$ subgroup of the original
U(1) survives. This describes the GSO projection on the current 
algebra fermions in the
heterotic string in the type I-heterotic duality context
\cite{pol-witt}.} 
In type I$^\prime$ theory, on the other hand, the projection is
$\Omega \I_9$. As a result the gauge field component along $S^1$
survives the projection. Similarily, while the NSNS two form $B$ is
odd under $\Omega$ and hence projected out in type I, $B_{\mu 9}$ is
also odd under $\I_9$ and thus survives the projection. This
guarantees the gauge invariance of the bound state.

As with most non-BPS configurations~\cite{sen,bg,gs}, the stability of
the $\Zop_2$ D-string, and its bound state with a fundamental
string, depends crucially on the size of the compact directions. We 
investigate this dependence by constructing boundary states which
describe the configurations studied presently. In particular we obtain
the critical radius at which the configurations become unstable and 
compute the tensions of these states. 

The boundary state representing a $\Zop_2$ D-string is given by
\be
\left|B1_{\Zop_2}\right>=\frac{{\cal N}}{2}
\left(\left|B1,+\right>_{\mbox{\scriptsize NSNS}}-
\left|B1,-\right>_{\mbox{\scriptsize NSNS}}\right)\,,
\ee
where ${\cal N}$ is the normalisation obtained by factorising on an
open string partition function (see for example~\cite{gs} for more
details on obtaining such normalisations) and
\ba
\ket{B1,k,\eta}_{NSNS}&=&\exp\left(\sum_{n=0}^\infty 
[\frac{1}{n}\alpha_{-n}^\mu S_{\mu\nu}\tilde{\alpha}_{-n}^\nu]
+i\eta\sum_{r\in\Nop+\frac{1}{2}} [\psi_{-r}^\mu
S_{\mu\nu}\tilde{\psi}_{-r}^\nu]\right)\nonumber \\
& &\;\;\;\;\;\;\times\;\;
\sum_{w_9}e^{i\theta w_9}
\ket{B1,k,w_9,\eta}^{(0)}
\otimes\ket{B1,\eta}_{\mbox{ghost}}\,.
\label{bdr}
\ea
Here $w_9$ is the winding number, $\eta=\pm 1$, $\theta$ is a Wilson line and 
for compactness of notation we do not write
$\theta$ on the left-hand side of the above equation. The matrix 
$S$ encodes the boundary conditions
of the D-string and is a $10\times 10$ diagonal matrix given by
\be
S=\mbox{diag}(1,1,\dots,1,-1)\,.
\ee
The D-string is taken to lie along directions $x^0$ and $x^9$ with
$x^9$ compactified along a circle of radius $R_{I^\prime}$, and we work in the
Minkowski metric. As the 
$\Zop_2$ D-string stretches in directions $0$ and $9$
while the O8-planes extend in directions $0,\dots,8$ it
is not possible to work in the light-cone gauge and ghost and
superghost contributions will have to be taken into account.
These are taken as in~\cite{clny,polcai,fetal}.
 The ground state $\ket{B1,k,w_9,\eta}^{(0)}$ carries momentum $k$
in the directions transverse to the D-string and is unique in the
NSNS sector. To obtain a localised D-brane, we have to Fourier
transform the above state,
\be
\ket{Bp,y,\eta}=\int\left(\prod_{\mu=1}^8dk^\mu e^{ik^\mu y_\mu}\right)
\sum_{w_9}e^{i\theta w_9}\ket{Bp,k,w_9,\eta}\,.
\ee
Here $y$ denotes the location of the boundary state along the
transverse directions, and is suppressed throughout as we take it to
be $y=0$.  The normalisation ${\cal N}$ is equal to that of a type IIB
D-string from which it follows that the tension of the $\Zop_2$
D-string is the same as that of a conventional type II
D-string. This is a 
$\sqrt{2}$ bigger than the tension of a type I BPS D-string. This
follows since the open string partition function for the $\Zop_2$
D-string has no GSO projection but does have the orientifold
projection $(1+\Omega\I_9)/2$.

Next consider the strings that end on a D8-brane. The boundary
state of a D8-brane is
\ba
\ket{B8}_{\mbox{\scriptsize NSNS}}&=&\frac{{\cal N}_8}{2}
(\ket{B8,+}_{\mbox{\scriptsize NSNS}}-\ket{B8,-}_{\mbox{\scriptsize NSNS}})
\nonumber \\
\ket{B8}_{\mbox{\scriptsize RR}}&=&\frac{4i{\cal N}_8}{2}
(\ket{B8,+}_{\mbox{\scriptsize RR}}+\ket{B8,-}_{\mbox{\scriptsize RR}})
\nonumber \\
\ket{B8}&=&\ket{B8}_{\mbox{\scriptsize NSNS}}+
\ket{B8}_{\mbox{\scriptsize RR}}\,.
\ea
Explicitly the NSNS boundary state is as in equation~(\ref{bdr}) but
now the matrix $S$ has eight entries equal to $-1$, with only the
first and last
one equal to 1. The RR sector is not discussed here as it does not
enter into our analysis. Factorisation of the cylinder diagram on an
annulus fixes the normalisation ${\cal N}_8$. 
Strings that stretch between a D8-brane and the $\Zop_2$ D-string 
have 9 ND and one NN boundary conditions. As a result the
ground state energy in the NS sector is positive, while as always in
the R sector it is zero, thus these are tachyon free.

Next we consider the M\"obius strip diagrams corresponding, in the
closed string channel to the
exchange between the D-string and the O8-planes. 
A crosscap state representing an O8-plane is given by a coherent 
state very similar to that of a D8-brane, the only difference 
being an extra factor of $(-1)^n$ in the coherent state 
exponentials~\cite{clny1,polcai}. To check that there are no open
string tachyons in the theory we compute the open string partition
functions. Since strings stretching 
between the D-string 
and the D8-brane have no tachyons we disregard these in the
following. The relevant amplitude in the closed string channel
is given by 
\ba
{\cal A}_1&=&\int dl\bra{B_{\Zop_2}1}e^{-lH_c}\ket{B_{\Zop_2}1}
+\sum_{i=1}^2\bra{B_{\Zop_2}1}e^{-lH_c}\ket{C_i8}
+\bra{C_i8}e^{-lH_c}\ket{B_{\Zop_2}1}\nonumber \\
&=&\frac{V}{2\pi}2^{-6}R_{I'}\int
dl\;l^{-4}\frac{f_3^8(q)-f_4^8(q)}{f_1^8(q)} 
\sum_{w_9=-\infty}^\infty e^{-l\pi(w_9R_{I'})^2}\nonumber \\
& &\;\;+\frac{V}{2\pi}4\times (-2^{-5/2})\int dl
\left(\frac{f_4^9(iq)f_1(iq)}{2^{-9/2}f_2^9(iq)f_3(iq)}-
\frac{f_3^9(iq)f_1(iq)}{2^{-9/2}f_2^9(iq)f_4(iq)}\right)\nonumber \\
&=&\frac{V}{2\pi}\frac{1}{2}\int \frac{dt}{2t} (2t)^{-1/2}
\frac{f_3^8({\tilde q})-f_2^8({\tilde q})}{f_1^8({\tilde q})}
\sum_{m=-\infty}^\infty e^{-2t\pi(m/R_{I'})^2}\nonumber \\
& &\;\;+\frac{V}{2\pi}2^{-3/2}\int \frac{dt}{2t}(2t)^{-1/2}
\left(e^{-i\pi/4}\frac{f_3^9(i{\tilde q})f_1(i{\tilde q})}
{2^{-9/2}f_2^9(i{\tilde q})f_4(i{\tilde q})}-
e^{i\pi/4}\frac{f_4^9(i{\tilde q})f_1(i{\tilde q})}
{2^{-9/2}f_2^9(i{\tilde q})f_3(i{\tilde q})}\right)\,,
\nonumber \\ \label{lpnf}
\ea
where $q=e^{-2\pi l}$, $t=1/2l$, ${\tilde q}=e^{-\pi t}$ and
$H_c$ is the closed string Hamiltonian 
\be
H_c=\pi p^2+2\pi\sum_{\mu=0}^9\left[\sum_{n=1}^\infty
(\alpha^\mu_{-n}\alpha^\mu_n+\tilde{\alpha}^\mu_{-n}\tilde{\alpha}^\mu_n)+
\sum_{r\in\Nop+\frac{1}{2}}
r(\psi^\mu_{-r}\psi^\mu_r+\tilde{\psi}^\mu_{-r}\tilde{\psi}^\mu_r)\right]
+2\pi C_c+\mbox{ghosts}\,.
\ee
The constant $C_c$ is $-1$. 

In order to see if there are open string tachyons we expand in
${\tilde q}$ to suitable order
\be
{\cal A}_1
=\frac{V}{2\pi}\half\int\frac{dt}{2t}(2t)^{-1/2}\left[\frac{1}{{\tilde
q}}(1+2{\tilde q}^{2/R^2})-\frac{1}{{\tilde q}}+\dots\right]\,.
\ee
It then follows that for $R_{I'} \le \sqrt{2}$,
the $\Zop_2$ D-string is tachyon free and hence stable. In
equation~(\ref{lpnf}) we have fixed the normalisation constants 
by requiring that the
cylinder and M\"obius strip diagrams factorise on suitable open string
partition functions.
The decay of the non-BPS D-string into a D0-$\overline{\mbox{D}0}$ pair
restricted to the orientifold planes is possible in the region 
\be
R_{I'} > \sqrt{2} \,. \label{RcI'1}
\ee
This is also confirmed by the fact that, in this region, the classical
mass of the D-string, given below, is bigger than that of two
fractional D-particles, given by~(\ref{mdf1}) and~(\ref{mdf2}), in
the above region
\be
m_{I'}(D1_{\mbox{\scriptsize nonbps}}) = {R_{I'} \over {\sqrt{2}
\lambda_{I'}}}\,. \label{mD1}  
\ee
The numerical factor above follows by noting that 
the tension of the $\Zop_2$ D-string is $\sqrt
2$ bigger than the tension of a type I BPS D-string. As expected,
the corresponding masses in the two theories are not related by the
duality map, since for non-BPS states the masses are not protected
from quantum corrections. In terms of heterotic string theory the
decay corresponds to~(\ref{v1v2}). The regimes of stability of the non-BPS
state in the two dual theories,~(\ref{Rchet1}) and~(\ref{RcI'1}), are
qualitatively the same, given the duality relation~(\ref{dmap}). 
   
Consider next a bound state of the $\Zop_2$ D-string with a
fundamental string. We show that the tension of this state is 
lower than the individual tenions of the D- and
F-strings, thus forming a bound state. Further we will analyse the
stability of this state and find a dependence on the radius. Boundary states 
with gauge fields were first considered
in~\cite{clny} and in the context of D-branes 
in~\cite{li,schmid,gg,fetal}. In the NSNS sector the
following changes occur in the boundary state description. 
The matrix $S$ now becomes
\be
S=\left(\begin{array}{cccc}\frac{\eta-F}{\eta+F}&&& \\ &-1&& 
\\&&\ddots& \\
&&&-1\end{array}\right)\,,
\ee
where $\eta$ is the Minkowski metric, there are still eight entries equal 
to $-1$ corresponding to the 
transverse directions and the field-strength $F$ is 
\be
F=\left(\begin{array}{cc}0&-f \\ f&0\end{array}\right)\,.
\ee
We have placed the $x^0$ and $x^9$ coordinates in positions 1,2 in
the matrix. In the above $f$ is given by \cite{fetal}
\be
f = -{m \over \sqrt{n^2 + m^2}}\,,
\ee
where $m = 1$, is the number of fundamental strings, and $n = 1$
is the number of D-strings. The background gauge field has no effect
on ghost or superghost contributions. In the presence of a gauge field
the open string momentum eigenvalues on a circle become~\cite{abou}
\be
p_n=\frac{n}{R}\frac{1}{\sqrt{1-f^2}}\,.\label{momg}
\ee
The non-compact open string momentum integrals get modified
in an analogous fashion. This can be viewed from the closed string
channel as an extra normalisation factor 
\be
\sqrt{-\det(\eta+F)}=\sqrt{1-f^2}\,,
\ee
of the boundary
state corresponding to the non-BPS bound state
relative to the $\Zop_2$ D-string~\cite{clny}. 
We are now in a position to compute the amplitude corresponding to
${\cal A}_1$ for the (F,D) bound state. Since $S$ is an
orthogonal matrix, the non-zero mode contributions to the
cylinder amplitude are as before. Further, the non-zero modes of the 
M\"obius strip amplitude do not change either as shown in the
Appendix. The amplitude is
\ba
{\cal A}_2&=&\int dl\bra{B1_{\Zop_2},F}e^{-lH_c}\ket{B1_{\Zop_2},F}
+\sum_{i=1}^2\bra{B1_{\Zop_2},F}e^{-lH_c}\ket{C8_i}
+\bra{C8_i}e^{-lH_c}\ket{B1_{\Zop_2},F}\nonumber \\
&=&\frac{V}{2\pi}2^{-6}R(1-f^2)
\int dl\;l^{-4}\frac{f_3^8(q)-f_4^8(q)}{f_1^8(q)}
\sum_{w_9=-\infty}^\infty e^{-l\pi(w_9R)^2(1-f^2)}\nonumber \\
& &\;+\frac{V}{2\pi}4\times (-2^{-5/2})\sqrt{1-f^2}\int dl
\left(\frac{f_4^9(iq)f_1(iq)}{2^{-9/2}f_2^9(iq)f_3(iq)}-
\frac{f_3^9(iq)f_1(iq)}{2^{-9/2}f_2^9(iq)f_4(iq)}\right)\nonumber \\
&=&\frac{V}{2\pi}\sqrt{1-f^2}\int \frac{dt}{2t} (2t)^{-1/2}
\frac{f_3^8({\tilde q})-f_2^8({\tilde q})}{f_1^8({\tilde q})}
\sum_{m=-\infty}^\infty e^{-2t\pi\left(\frac{m}{R\sqrt{1-f^2}}
\right)^2}\nonumber \\
& &\;+\frac{V}{2\pi}\sqrt{1-f^2} 
2^{-3/2}\!\int \!\frac{dt}{2t^2}(2t)^{-1/2}
\left(e^{-i\pi/4}\frac{f_3^9(i{\tilde q})f_1(i{\tilde q})}
{2^{-9/2}f_2^9(i{\tilde q})f_4(i{\tilde q})}-
e^{i\pi/4}\frac{f_4^9(i{\tilde q})f_1(i{\tilde q})}
{2^{-9/2}f_2^9(i{\tilde q})f_3(i{\tilde q})}\right)\,,
\nonumber \\\label{fin}
\ea
which we again expand to suitable order
\be
{\cal A}_2=\frac{V}{2\pi}\half\sqrt{1-f^2}
\int\frac{dt}{2t}(2t)^{-1/2}\left[\frac{1}{{\tilde
q}}(1+2{\tilde q}^{2/(R^2(1-f^2))})-2^{-1/2}\frac{1}{{\tilde q}}
+\dots\right]\,.
\ee
It then follows that for $R_{I'} \le \sqrt{\frac{2}{1-f^2}} = 2$ the 
(F,D) bound state is stable and tachyon free. Hence the decay of the
non-BPS (F,D) bound state into a D0-$\overline{\mbox{D}0}$ pair and a
fundamental string is possible in the region
\be
R_{I'} > 2\,. \label{RcI'2}
\ee
Again the regimes of stability of the non-BPS
state in the two dual theories,~(\ref{Rchet2}) and~(\ref{RcI'2}), are
qualitatively 
the same, given the duality relation~(\ref{dmap}). 
The tension of a general configuration of $m$ F-string and $n$
D-string bound state is $(m^2 + n^2)^{1/2}$ times the D-string
tension \cite{fetal}. Non-threshold bound states are realised if $m$
and $n$ are relatively prime integers. In our case the non-BPS
D-string mass is given by~(\ref{mD1}). Hence the mass of the non-BPS 
D-string with one unit of electric field (with $m=n=1$) is given by 
\be
m_{I'} ( D1_{\mbox{\scriptsize nonbps}} + F) 
= {R_{I'} \over \lambda_{I'}}. \label{mDE} 
\ee
Unlike the case of the non-BPS D-string, the mass of the non-BPS
(F,D) bound state is not same as its decay
products (namely the D0-$\overline{\mbox{D}0}$ pair with (8,8) F-string) at 
the critical radius~(\ref{RcI'2}). Here the F-string mass is
$R_{I'}/2$ 
obtained from~(\ref{mbv}) using duality map~(\ref{dmap}). There is a
loss of energy due to the interaction between the decay products in the 
presence of an electric
field. The transition from the non-BPS bound state to the D-particle
pair and a fundamental string at the critical radius does not turn out to be
a marginal deformation.  
\section{Type I analysis}
\setcounter{equation}{0}
In this section we briefly outline what happens under T-duality to the
analysis of the previous section. As expected T-duality remains valid
when analysing non-BPS states. Under T-duality the $\Zop_2$
D-string becomes a $\Zop_2$ D-particle analysed in~\cite{sen4,fraunbps}. 
The combined cylinder and M\"obius
strip diagram amplitude~(\ref{lpnf}) undergoes only one change; the
open string momentum sum $\sum_m e^{-2t(m/R_{I^\prime})^2}$ now becomes
a winding sum $\sum_w e^{-2t(wR_I)^2}$, thus stabilising the D-particle for 
\be
R_{I}>\frac{1}{\sqrt{2}}\,,
\ee
in agreement with~\cite{sen4,fraunbps}.
Under T-duality an electric field becomes a velocity, and so the
$\Zop_2$ D-string with electric flux is dual to a D-particle with 
constant velocity in the compactified direction. The open string momentum sum
in~(\ref{fin}) now becomes a winding sum
\be
\sum_w e^{-2t(wR_I\sqrt{1-f^2})^2}\,.
\ee
The $f$ dependent square-root is easily identified as a Lorentz
contraction of the compactification radius as viewed by the moving
$\Zop_2$ D-particle. This moving D-particle is stable for
\be
R_{I}>\frac{1}{2}\,,
\ee
a radius smaller than that of the static D-particle, which can be
regarded as a consequence of Special Relativity.
\section{Final Remarks and Conclusion}
Let $(p,q)$ denotes a bound state of $p$ F-strings and
$q$ D-strings with $p,q$ any relatively prime pair in type IIB
string theory. Existence of this bound state is a consequence of
conjectured $SL(2,\Zop)$ duality symmetry of the
type IIB string theory~\cite{schw}. The $(p,1)$ bound states
exists as a consequence of the D-string structure \cite{witt1}. 
But the situation is different in our case with $\Zop_2$-charged
non-BPS D-string due to the following observation.  
From equation~(\ref{bsbv}), we see that if we add {\it even} number of
bi-vector states to any bi-spinor in ${\cal A}$ we get a bi-spinor in
${\cal A}$
itself. In type I$^\prime$ theory, this observation corresponds to the fact 
that $( p,1 )$ bound states exist only for $p=1$. For $p$
odd, the bound state decays into a $p=1$ bound state and a number of
winding states. On the other hand, 
if we add two (8,8) F-strings to the non-BPS D-string the 
system is unstable against decay into the D-string itself and
(probably) a full winding state or a closed string state.     
As far as charge conservation is concerned this is equivalent to the
fact that two such D-strings together are unstable and decay into
massless states. The fact that two D-strings in our example are
not stable is not surprising as two spinor states can annihilate to
give various massless states.
For example, if we take two bi-spinor states of the form $((
{1 \over 2})^8 ;( {1 \over 2})^8)$ they
together carry the same charge and same mass (at all radii) as the
state $\left( 1^{16} \right)$ which can decay into various massless
states in the adjoint representation describing (8,8) F-strings with
both the D8-branes on the same orientifold planes. 

Since the mass of a state with unit winding is $R_{I'}$,
in the case of the D-string with two units of electric field,
using~(\ref{mDE}), the inequality  
\be
m_{I'} (\mbox{D-string} + 2F ) \,\,>\,\, 
m_{I'} (\mbox{D-string} ) + m_{I'} (\mbox{winding state} )\,,
\ee
holds for sufficiently small $\lambda_{I'}$ at all radius. Hence, for a
sufficiently small type I$^\prime$ coupling, the D-string with two
units of electric field is unstable against a decay into a D-string and a
closed string state. Since the configuration studied is non-BPS, it is
not surprising that the above inequality only holds for small type
I$^\prime$ coupling. 

In this paper we have tested the duality between the heterotic and
type I$^\prime$ string with $SO(16) \times SO(16)$ gauge group beyond
the BPS limit. We have found agreement between the regions of
stability and decay products of non-BPS states in both theories.
 In particular we have found that the domains of stability are
qualitativety related by the duality map between the two
theories. This fact was not guaranteed {\em a priori} as the masses of
non-BPS states are not protected by supersymmetry. Unlike the case of
non-BPS D-string the mass of the non-BPS (F,D) bound state is not the
same as its decay product at the critical radius. This indicates that
the transition from the non-BPS bound state to the D-particle pair
and a fundamental string is not a marginal deformation.   

It would be interesting to perform a similar analysis for other Wilson
lines, where th gauge group is $SO(16-2N)
\times SO(16+2N)$. This is achieved by  moving $N$ D8-branes from
one O8-plane to the other and to see whether there is any possible
modification in string creation and gauge enhancement phenomena in
the presence of these non-BPS D-branes. 

\appendix

\section{M\"obius Strip Diagram}
\setcounter{equation}{0}

In this appendix, we explicitly derive the M\"obius strip part of 
equation~(\ref{fin}). The term we are then interested in is
\be
{\cal M}=\bra{C8}e^{-lH_c}\ket{B1_{\Zop_2},F}\,.
\ee
The contributions from the ghosts, superghosts and directions 
transverse to the $\Zop_2$ D-string are as in the case of $F=0$. 
We focus here on the matter contributions in the directions $x^0$ 
and $x^9$. We write
\be
\frac{\eta-F}{\eta+F}=\left(\begin{array}{ll}\cosh(2\nu)& \sinh(2\nu) \\
\sinh(2\nu) & \cosh(2\nu)\end{array}\right)\,,
\ee
since the matrix is orthogonal relative to the inner product defined
by the Minkowski metric $\eta$. The bosonic contribution to ${\cal M}$
is given by
\ba
& &\bra{C8}e^{-lH_c}\ket{B_{\Zop_2}1,F}_{0,9}\nonumber \\
&=&\bra{0}
e^{\sum_{m=1}^\infty\frac{(-1)^m}{m}\left(\alpha^0_m{\tilde\alpha}^0_m
+\alpha^9_m{\tilde\alpha}^9_m\right)}e^{-lH_c}\nonumber \\
& &\times\;\;
e^{\sum_{n=1}^\infty\frac{1}{n}\left(
\alpha^0_{-n}({\tilde\alpha}^0_{-n}\cosh(2\nu)+{\tilde\alpha}^9_{-n} 
\sinh(2\nu))
-\alpha^9_{-n}({\tilde\alpha}^9_{-n}\cosh(2\nu)+{\tilde\alpha}^0_{-n} 
\sinh(2\nu))\right)}\ket{0}\nonumber \\
&=&\prod_{n=1}^\infty\sum_{k,l,j,p,s,t=0}^\infty\bra{0}
\frac{\left(\frac{(-1)^n}{n}\alpha^0_n{\tilde\alpha^0_n}\right)^k}{k!}
\frac{\left(\frac{(-1)^n}{n}\alpha^9_n{\tilde\alpha^9_n}\right)^l}{l!}
\frac{\left(\frac{1}{n}\alpha^0_{-n}q^{2n}
{\tilde\alpha}^0_{-n}\cosh(2\nu)\right)^j}{j!}
\nonumber \\
& &\times\;\;
\frac{\left(-\frac{1}{n}\alpha^9_{-n}q^{2n}
{\tilde\alpha}^9_{-n}\cosh(2\nu)\right)^p}{p!}
\frac{\left(\frac{1}{n}\alpha^0_{-n}q^{2n}
{\tilde\alpha}^9_{-n}\sinh(2\nu)\right)^s}{s!}
\frac{\left(-\frac{1}{n}\alpha^9_{-n}q^{2n}
{\tilde\alpha}^0_{-n}\sinh(2\nu)\right)^t}{t!}\ket{0} \nonumber \\
&=&\prod_n\sum_{k,l,j}\bra{0}
\frac{\left(\frac{(-1)^n}{n}\alpha^0_n{\tilde\alpha^0_n}\right)^{k+j}}
{(k+j)!}
\frac{\left(\frac{1}{n}\alpha^0_{-n}q^{2n}
{\tilde\alpha}^0_{-n}\cosh(2\nu)\right)^k}{k!}
\frac{\left(\frac{(-1)^n}{n}\alpha^9_n{\tilde\alpha^9_n}\right)^{l+j}}
{(l+j)!}
\nonumber \\
& &\times\;\;
\frac{\left(-\frac{1}{n}\alpha^9_{-n}q^{2n}
{\tilde\alpha}^9_{-n}\cosh(2\nu)\right)^l}{l!} 
\frac{\left(-\frac{1}{n^2}q^{4n}\alpha^0_{-n}{\tilde\alpha}^0_{-n}
\alpha^9_{-n}{\tilde\alpha}^9_{-n}\sinh^2(2\nu)\right)^j}{j!j!}
\ket{0}\nonumber\\
&=&\prod_n
\sum_{k,l,j}\frac{(k+j)!(l+j)!(iq)^{2n(k+l+2j)}\cosh^{k+l}(2\nu)
\sinh^{2j}(2\nu)}{k!l!n^{4j}(j!)^6}\nonumber \\
& &\times\;\;
\bra{0}\left(\alpha^0_n{\tilde\alpha}^0_n\alpha^9_n{\tilde\alpha}^9_n
\right)^j
\left(-\alpha^0_{-n}{\tilde\alpha}^0_{-n}\alpha^9_{-n}{\tilde\alpha}^9_{-n}\right)^j\ket{0}\nonumber \\
&=&\prod_n
\sum_{k,l,j}\left(\begin {array}{c}k+j \\j \end{array}\right)
\left(\begin {array}{c}l+j \\j \end{array}\right)
(iq)^{2n(k+l+2j)}\cosh^{k+l}(2\nu)(-\sinh(2\nu))^{2j}\nonumber \\
&=&\sum_j(1+(iq)^{2n}\cosh(2\nu))^{-j-1}(1-(iq)^{2n}\cosh(2\nu))^{-j-1}((iq)^{4n}\sinh^2(2\nu))^j\nonumber \\
&=&\prod_n(1-(iq)^{2n})^{-1}(1+(iq)^{2n})^{-1}\,.
\ea
In the above we have used the following 
\be
\sum_{k=0}^\infty
\left(\begin {array}{c}k+j \\j \end{array}\right)a^k
=(1-a)^{-j-1}\,.
\ee
This demonstrates that the bosonic non-zero modes contribute the same
factor to the M\"obius strip amplitude with or without a gauge field. The
fermions are substantially easier as the coherent state exponentials
terminate, due to the anti-commutative nature of the $\psi^\mu_r$.
Explicitly we have
\ba
& &\bra{C8}e^{-lH_c}\ket{B_{\Zop_2}1,F}_{0,9}\nonumber \\
&=&\bra{0}
e^{\sum_{r\in\Nop+\frac{1}{2}}(-1)^ri\left(\psi^0_r{\tilde\psi}^0_r
+\psi^9_r{\tilde\psi}^9_r\right)}e^{-lH_c}\nonumber \\
& &\times\;\;
e^{\sum_{s\in\Nop+\frac{1}{2}}i\left(
\psi^0_{-s}({\tilde\psi}^0_{-s}\cosh(2\nu)+{\tilde\psi}^9_{-s} 
\sinh(2\nu))
-\psi^9_{-s}({\tilde\psi}^9_{-s}\cosh(2\nu)+{\tilde\psi}^0_{-s} 
\sinh(2\nu))\right)}\ket{0}\nonumber \\
&=&\prod_{r\in\Nop+\half}\bra{0}(1+i(-1)^r\psi^0_r{\tilde\psi}^0_r)
(1+i(-1)^r\psi^9_r{\tilde\psi}^9_r)(1+iq^{2r}\cosh(2\nu)
\psi^0_r{\tilde\psi}^0_r)\nonumber \\ & &\times \;\;
(1-iq^{2r}\cosh(2\nu)\psi^9_r{\tilde\psi}^9_r)
(1+iq^{2r}\sinh(2\nu)\psi^0_r{\tilde\psi}^9_r)
(1-iq^{2r}\sinh(2\nu)\psi^9_r{\tilde\psi}^0_r)\ket{0}\nonumber \\
&=&\prod_{r\in\Nop+\half}
\bra{0}(1+i(-1)^r\psi^0_r{\tilde\psi}^0_r+i(-1)^r\psi^9_r{\tilde\psi}^9_r
-(-1)^{2r}\psi^0_r{\tilde\psi}^0_r\psi^9_r{\tilde\psi}^9_r)
\nonumber \\& &\times\;\;
(1+iq^{2r}\cosh(2\nu)\psi^0_r{\tilde\psi}^0_r
-iq^{2r}\cosh(2\nu)\psi^9_r{\tilde\psi}^9_r
+q^{4r}\cosh^2(2\nu)\psi^0_r{\tilde\psi}^0_r
\psi^9_r{\tilde\psi}^9_r)\nonumber \\ & &\times\;\;
(1+q^{4r}\sinh^2(2\nu)\psi^0_r{\tilde\psi}^9_r
\psi^9_r{\tilde\psi}^0_r)\ket{0}\nonumber \\
&=&\prod_{r\in\Nop+\half}
\left(1-(iq)^{4r}(cosh^2(2\nu)-sinh^2(2\nu))\right)\nonumber \\
&=&\prod_{n=1}^\infty(1-(iq)^{2n-1})(1+(iq)^{2n-1})\,,
\ea
which is indeed the same as for the case with no gauge field.

\section*{Acknowledgements}

We are grateful to M.B. Green, A. Sen, M.C. Daflon Barrozo for many
useful discussion and in particular to M.R. Gaberdiel for support
throughout the project.\\
T.D. is supported by a Trinity College Scholarship.
B.S. is supported by the Cambridge Overseas Trust.

\end{document}